\documentclass[acmsmall,nonacm]{acmart}
 \usepackage{hyperref}
	\usepackage{color,soul}
 \usepackage{graphicx}
	\usepackage{diagbox}
	\usepackage{makecell}
\usepackage[figuresright]{rotating}
\usepackage{lscape}
	\usepackage{multirow,multicol}
	\usepackage{textcomp}
 \usepackage{setspace}
	\usepackage{enumitem}
 \usepackage{outlines}
	\usepackage{multicol}
    \usepackage{multirow}
    \usepackage{tabularx}
    \usepackage{makecell}
    \usepackage{colortbl}
    \usepackage{caption}
    \usepackage{subfigure}
    \usepackage{lipsum}
    \usepackage[utf8]{inputenc}
    \usepackage{algorithm2e}
    \usepackage{array}
    \usepackage{xcolor}
    \usepackage[flushleft]{threeparttable}

\newcommand{\gray}[1]{\textcolor{gray}{#1}}

\newcommand{\PreserveBackslash}[1]{\let\temp=\\#1\let\\=\temp}
\newcolumntype{C}[1]{>{\PreserveBackslash\centering}p{#1}}

\AtBeginDocument{%
  \providecommand\BibTeX{{%
    \normalfont B\kern-0.5em{\scshape i\kern-0.25em b}\kern-0.8em\TeX}}}

\begin{document}

\title[CommSense: A Computational Framework for Evaluating Patient-Provider Interactions]{CommSense: A Wearable Sensing Computational Framework for Evaluating Patient-Clinician Interactions}


\author{Zhiyuan Wang}
\authornote{Corresponding author: Zhiyuan Wang (vmf9pr@virginia.edu)}
\email{vmf9pr@virginia.edu}
\affiliation{%
  \institution{Department of Systems and Information Engineering, University of Virginia}
  \streetaddress{151 Engineer's Way}
  \city{Charlottesville}
  \state{Virginia}
  \country{USA}
  \postcode{22904}
}

\author{Nusayer Hassan}
\email{sh2mt@virginia.edu}
\affiliation{%
  \institution{Department of Systems and Information Engineering, University of Virginia}
  \streetaddress{151 Engineer's Way}
  \city{Charlottesville}
  \state{Virginia}
  \country{USA}
  \postcode{22904}
}

\author{Virginia LeBaron}
\email{vtl6k@virginia.edu}
\affiliation{%
  \institution{School of Nursing, University of Virginia}
  \streetaddress{225 Jeanette Lancaster Way}
  \city{Charlottesville}
  \state{Virginia}
  \country{USA}
  \postcode{22903}
}

\author{Tabor E. Flickinger}
\email{tes3j@virginia.edu}
\affiliation{%
  \institution{School of Medicine, University of Virginia}
  \streetaddress{1340 Jefferson Park Ave}
  \city{Charlottesville}
  \state{Virginia}
  \country{USA}
  \postcode{22903}
}

\author{David Ling}
\email{dyl3a@uvahealth.org}
\affiliation{%
  \institution{School of Medicine, University of Virginia}
  \streetaddress{1340 Jefferson Park Ave}
  \city{Charlottesville}
  \state{Virginia}
  \country{USA}
  \postcode{22903}
}

\author{James Edwards}
\email{jyq2ey@virginia.edu}
\affiliation{%
  \institution{School of Nursing, University of Virginia}
  \streetaddress{225 Jeanette Lancaster Way}
  \city{Charlottesville}
  \state{Virginia}
  \country{USA}
  \postcode{22903}
}

\author{Congyu Wu}
\email{congyu.wu@binghamton.edu}
\affiliation{%
  \institution{Department of Systems Science and Industrial Engineering, Binghamton University}
  \streetaddress{Engineering Bldg, L2}
  \city{Binghamton}
  \state{New York}
  \country{USA}
  \postcode{13902}
}

\author{Mehdi Boukhechba}
\email{mboukhec@its.jnj.com}
\affiliation{%
  \institution{Johnson \& Johnson Innovative Medicine}
  \city{Titusville}
  \state{New Jersey}
  \country{USA}
  \postcode{08560}
}

\author{Laura E. Barnes}
\email{lb3dp@virginia.edu}
\affiliation{%
  \institution{Department of Systems and Information Engineering, University of Virginia}
  \streetaddress{151 Engineer's Way}
  \city{Charlottesville}
  \state{Virginia}
  \country{USA}
  \postcode{22904}
}

\renewcommand{\shortauthors}{Wang, et al.}

\begin{abstract}

    Quality patient-provider communication is critical to improve clinical care and patient outcomes. While progress has been made with communication skills training for clinicians, significant gaps exist in how to best monitor, measure, and evaluate the implementation of communication skills in the actual clinical setting. Advancements in ubiquitous technology and natural language processing make it possible to realize more objective, real-time assessment of clinical interactions and in turn provide more timely feedback to clinicians about their communication effectiveness. In this paper, we propose CommSense, a computational sensing framework that combines smartwatch audio and transcripts with natural language processing methods to measure selected ``best-practice'' communication metrics captured by wearable devices in the context of palliative care interactions, including understanding, empathy, presence, emotion, and clarity. We conducted a pilot study involving N=40 clinician participants, to test the technical feasibility and acceptability of CommSense in a simulated clinical setting. Our findings demonstrate that CommSense effectively captures most communication metrics and is well-received by both practicing clinicians and student trainees. Our study also highlights the potential for digital technology to enhance communication skills training for healthcare providers and students, ultimately resulting in more equitable delivery of healthcare and accessible, lower cost tools for training with the potential to improve patient outcomes. 
 \end{abstract}



\maketitle

\section{Introduction}

Effective clinical communication is critical for healthcare providers to provide high-quality care, establish rapport and trust with patients, and enhance patient outcomes \cite{parker2000improving,curtis2013effect}. The interactions encompass not only the exchange of information but also the conveyance of empathy, understanding, and emotional support \cite{halpern2003clinical,geerse2019qualitative}. In the context of palliative care (care provided to patients and their families coping with serious, life-limiting illness) \cite{morrison2004palliative}, the requirements of effective communication become even more nuanced and complex, requiring a heightened level of empathy and emotional understanding \cite{goldsmith2013palliative}. Ineffective communication can lead to significant negative emotional, physical, and even financial consequences for patients and family caregivers \cite{thorne2008cancer}. Healthcare organizations have therefore prioritized communication skills education and training for healthcare trainees and providers \cite{burns2012training}. More notably, poor communication in healthcare can exacerbate systemic issues within the landscape of health equity, especially for marginalized racial and ethnic groups. Empirical evidence illustrates the minority groups' heightened risks of experiencing undertreated pain \cite{anderson2009racial}, dying in intensive care units against their expressed preferences \cite{elliott2016differences,mack2010racial}, and encountering subpar communication regarding their healthcare needs \cite{elliott2016differences,palmer2014racial,long2014race}. Thus, these communication disparities act as a formidable obstacle to health equity, emphasizing the urgent need for effective and scalable solutions.

However, current methods to evaluate healthcare communication performance in clinical settings, provide actionable feedback, and track progress longitudinally are limited \cite{tulsky2017research,brighton2017systematic,sanders2020quality}. Specifically, existing methods are often reliant on a limited number of human coders, an approach that is not only expensive but also susceptible to human biases \cite{byerly1969nurse,sennekamp2012development}. Additionally, despite robust research efforts (e.g., VitalTalk \cite{arnold2017oncotalk}) on providing feedback on clinical conversations, achieving real-world scalability, adaptability, and longitudinal progress tracking in healthcare communication remains significant challenges \cite{tsimtsiou2017enhancing,sanders2020quality,hovland2023patients}. Given these challenges, there is a clear gap in the provision of objective, continuous, and fine-grained assessments of communication skills. These measures should be designed with the dual goal of providing feedback and facilitating improvement in clinicians' communication skills.

Recently, advancements in ubiquitous and mobile sensing \cite{lane2010survey,wang2022personalized} and natural language processing (NLP) techniques \cite{hirschberg2015advances} have shown promise for more adaptive and automated measures of communication effectiveness \cite{lindvall2019natural,lee2020natural,lebaron2022exploring,flickinger2023evidence}. The combination of these advancements in sensing technologies has yielded interactive mobile systems and embedded sensors that, when paired with computational methods such as NLP, can offer innovative approaches to communication assessment, paving the way for substantial improvement in palliative care \cite{chen2018bibliometric}. While existing programs like COMFORT\footnote{The COMFORT Communication Project. \url{https://www.communicatecomfort.com/}} \cite{fuoto2019palliative} have made progress in enhancing healthcare communication, a significant gap persists in developing a scalable solution that integrates \textbf{ubiquitous wearable sensors and NLP technology for the evaluation of communication in real clinical settings}.

In this paper, we propose CommSense, a computational sensing framework that can leverage wearable sensing technology to assess and enhance patient-provider communication. Firstly, through a literature review, we identify the ``best-practice'' communication performance metrics to be captured and extracted by wearable devices in the context of palliative care, including understanding, empathy, presence, emotion, and clarity. Through a combination of ubiquitous sensing and NLP technologies, this framework captures and analyzes audio and speech from patient-provider interactions. With self-trained deep language models, pre-trained acoustic models, and word-use and semantic analyses, the data is computed under designed operationalization rules to provide assessment of the communication metrics and personalized feedback to healthcare providers, facilitating continuous improvement in their communication skills. We pilot tested the technical feasibility and capability of the proposed methods with N=40 participants, consisting of clinicians, nursing students, and medical students, in a simulated clinical setting. The evaluation outcomes revealed promising results with an average over 77.9\%, 71.2\%, 70.2\% in balanced accuracy, precision, and recall, respectively, in identifying the manually-labeled ground truth (good or bad) of each communication metric in the interaction segment. The findings setting the preliminary stage for future work and testing in real-world clinical contexts.

This study makes the following contributions:

\begin{itemize}
    \item We introduce, to our knowledge, the first digital technology prototype that leverages mobile and ubiquitous sensing alongside state-of-the-art NLP techniques to assess patient-clinician interactions in the context of palliative care. This prototype aims to demonstrate the feasibility of automating patient-provider communication pattern monitoring and assessment moving towards more objective, timely, longitudinal assessment of patient-clinician interactions with potential to improve patient satisfaction and outcomes.
    \item To actualize this vision, we propose a computational sensing framework, CommSense. This framework focuses on identifying core `best-practice' communication metrics using integrated computational approaches. Leveraging advanced linguistic and NLP techniques, we automate the evaluation of these communication metrics.
    \item We validate CommSense through a feasibility pilot with forty clinical participants using simulated recordings collected. Achieving an average accuracy exceeding 77.9\%, this work underscores the potential of our digital wearable approach for enhancing healthcare communication and sets the stage for future NLP efforts to further improve the extraction of specific communication metrics, such as detecting the proficiency of providers in terms of presence and empathy.
\end{itemize}

\section{Related Work}

There are a multitude of studies that highlight the importance of effective patient-clinician communication in healthcare settings \cite{parker2000improving,curtis2013effect,halpern2003clinical,geerse2019qualitative}. These studies underscore the need for a automated and scalable mechanism to assess, improve, and maintain the communication skills of healthcare providers. In the following, we provide a detailed synthesis of pivotal literature in three domains: 1) CSCW in patient-provider interactions, 2) clinical evaluation strategies and techniques, and 3) role of technology in evaluation of clinical communication.

\subsection{CSCW in Patient-Provider Interactions} \label{sec:literature_cscw}

Computer-supported cooperative work (CSCW) has long been recognized for its crucial role in improving patient-provider interactions \cite{fitzpatrick2013review,pichon2021divided}. Key studies have demonstrated how CSCW principles can be applied to develop systems that facilitate effective communication between patients and providers \cite{reddy2001coordinating, unertl2009describing}. For instance, Reddy \textit{et al.} \cite{reddy2001coordinating} underscored how CSCW could be harnessed to coordinate a variety of tasks in medical care, thereby making healthcare services more patient-centered. Extending this notion, Unertl \textit{et al.} \cite{unertl2009describing} illustrated how CSCW could aid in modeling workflow and information flow in chronic disease care, effectively enhancing healthcare delivery. The advancement of Information and Communication Technologies (ICTs) has also significantly impacted patient-provider communication. Studies such as Chen \textit{et al.} \cite{chen2011unpacking} addressed the challenge of how computer presence in the consultation room could potentially detract from patient engagement. They suggested solutions like "Computer-on-Wheels" (COWs), which could be reoriented and repositioned during different phases of a medical visit, thereby facilitating better patient participation and patient-physician eye contact. Additionally, Ding \textit{et al.} \cite{ding2019boundary} ventured into the personal utilization of ICTs for patient-provider communication. They found that mobile social media application could strengthen patient-provider relationships and provide psychological reassurance.

Despite these technological development, a significant gap still exists in capturing the complexity of patient-provider interactions, particularly the elements of empathy, understanding, and emotional support. These elements are not arbitrary but recognized as crucial components of effective clinical communication, impacting patient engagement, adherence to treatment, and overall satisfaction \cite{zolnierek2009physician,riess2010empathy}. This gap is accentuated in complex chronic conditions, such as metastatic cancer, where the need for patients and providers to find common ground and align perspectives, vital for shared decision-making, is crucial \cite{butow2002communicating,elwyn2012shared}. Therefore, there remains a pressing need for further research and development in systems designed to assess and improve the quality of these interactions.

\subsection{Clinical Communication Evaluation Strategies and Techniques}

Conventional clinical communication assessment has predominantly depended on human observers and subjective evaluations, often utilizing structured observational methods such as the Roter Interaction Analysis System (RIAS) \cite{roter2002roter} and the Calgary-Cambridge guide \cite{kurtz2003marrying}. While these methods have been instrumental in understanding structured communication and facilitating teaching and learning, their validity frequently suffers from difficulties in widely applied and inter-rater discrepancies \cite{price2008assessing}. Such human-observer based rating systems come with limitations: they involve substantial labor and documentation cost and burden, difficult for widescale implementation, and are prone to disparities and human biases (e.g., recall bias, cognitive bias, and confirmation bias) \cite{schacter2002seven,nickerson1998confirmation}. However, wearable sensing and NLP in human research also encounter inherent biases in data gathering (e.g., lack of training data from non-English speakers), algorithmic processing (algorithmic biases), and interpretation \cite{hemmer2019let,offenwanger2021diagnosing,10.1145/3334480.3382892}.

Another approach relies on patient feedback through rating scales such as the Consultation and Relational Empathy (CARE) Measure \cite{mercer2005relevance} and AIDET survey \cite{panchuay2023exploring}, where the patient provides feedback to the clinician about their communication performance. However, this method is clearly subject to patient bias and can be influenced by factors unrelated to communication skills \cite{bikker2015measuring}, leading to a potential failure in objectively assessing the technical proficiency of healthcare providers. Additionally, self-assessment (e.g., self-review by watching video replay) is an alternative that has been used to evaluate clinical communication, which allows for reflection and self-improvement \cite{hawkins2012improving}. However, it is well-documented that healthcare professionals often have inaccurate perceptions of their skills, which can lead to over- or under-estimation of their communication capabilities \cite{davis2006accuracy}.

Recently, efforts have been made to utilize technology to gather interaction data and practice communication skills. These include the use of simulation, virtual patients \cite{cook2010computerized,datta2016deep}, and online meetings \cite{thampy2022virtual}. While promising, the practicality of these techniques in naturalistic settings remains unproven. Given that most traditional methods of assessment are resource-intensive, with potential bias, lack objectivity, and do not lend themselves to widely accessible and longitudinal assessment, ubiquitous technology presents a unique opportunity to improve the automated assessment of provider communication with patients. By leveraging ubiquitous sensing and advanced NLP, CommSense has the potential to provide a more comprehensive and immediate evaluation and personalized metrics for communication assessment, marking a significant step towards improved clinical communication measurement and, in turn, reduced health disparities.

\subsection{Role of Technology in Clinical Communication} \label{sec:literature_tech}

The transformative role of technology in healthcare communication can be summarized under two broad categories: ubiquitous sensing systems and NLP computational methods.

\subsubsection{Ubiquitous Sensing Systems}

Ubiquitous sensing systems, including wearables, portable devices, wireless communication mechanisms, and multimedia tools, are explored to be applied to monitor, capture, and analyze clinical interactions and patient information \cite{ko2010wireless,deen2015information,lebaron2022deploying,hovland2023patients}. These systems offer unique opportunities for objective observation of healthcare interactions, overcoming the biases and limitations inherent in human-based observation.

Early initiatives have utilized these systems for evaluating clinical behavior. Notable examples include a portable sociometric badge system that tracks speech activity and interactions among healthcare professionals and wearable devices that offer insightful results in monitoring clinical activities, such as bedside nursing care \cite{kim2008meeting,olguin2009wearable}. The application of wireless technology in healthcare communication has also seen significant advancements \cite{hovland2023patients}.

Sensing systems based on multimedia (e.g., audio and video) provide another dimension, improving documentation and patient understanding by deploying voice recognition technology for transcribing patient-provider conversations \cite{johnson2014systematic}. Research has also delved into body language sensing systems to decode physical and mental states and strengthen patient-provider rapport \cite{haque2020illuminating}. However, there is an evident gap in research on using these systems to understand and enhance verbal communication, a challenge that CommSense is designed to address.

\subsubsection{NLP Computational Methods} 

Beyond sensing systems, computational methodologies offer powerful tools for understanding and improving healthcare communication. NLP techniques have been deployed to identify communication barriers, anticipate patient outcomes, and evaluate communication skills \cite{lindvall2019natural,lee2020natural}.

Several studies have demonstrated implementation of NLP in clinical contexts. For example, Townsend et al. \cite{townsend2013natural} emphasized the potential of NLP to significantly enhance clinical outcomes, ranging from transcribing text reports to analyzing electronic health records. Durieux et al. \cite{durieux2023development} developed a keyword library for detecting symptom talk in oncology conversations, highlighting how computational methods can help attend to patient suffering more effectively. Ross et al. \cite{ross2020story} utilized NLP to analyze complex clinical narratives from serious illness conversations, revealing the hidden lexicon within these narratives and provided NLP insights into understanding intricate clinical conversations. Similarly, Bhatt et al. \cite{bhatt2023use} used NLP to assess social support in patients with advanced cancer, evaluating important social factors in healthcare and revealing possible relationships between social support and healthcare utilization.

Recently, we have seen increasing integration of artificial intelligence systems with NLP techniques. For example, Zhang et al. \cite{zhang2022conversational} proposed an AI-powered conversational system for clinical communication training, showcasing the potential of NLP in adapting communication training methods. Ali et al. \cite{ali2023using} reported the early adoption of ChatGPT to generate patient clinic letters, then to understand clinical communication. 

In conclusion, while there have been promising advancements in utilizing NLP for clinical communication, the domain still has a significant gap to fill by CommSense – the development of systems that can automatically evaluate the quality of patient-provider communication during real-time clinical encounters and provide actionable feedback to clinicians.

\section{CommSense: A Computational Sensing Framework}

\begin{figure}[t!]
\includegraphics[width=1\columnwidth]{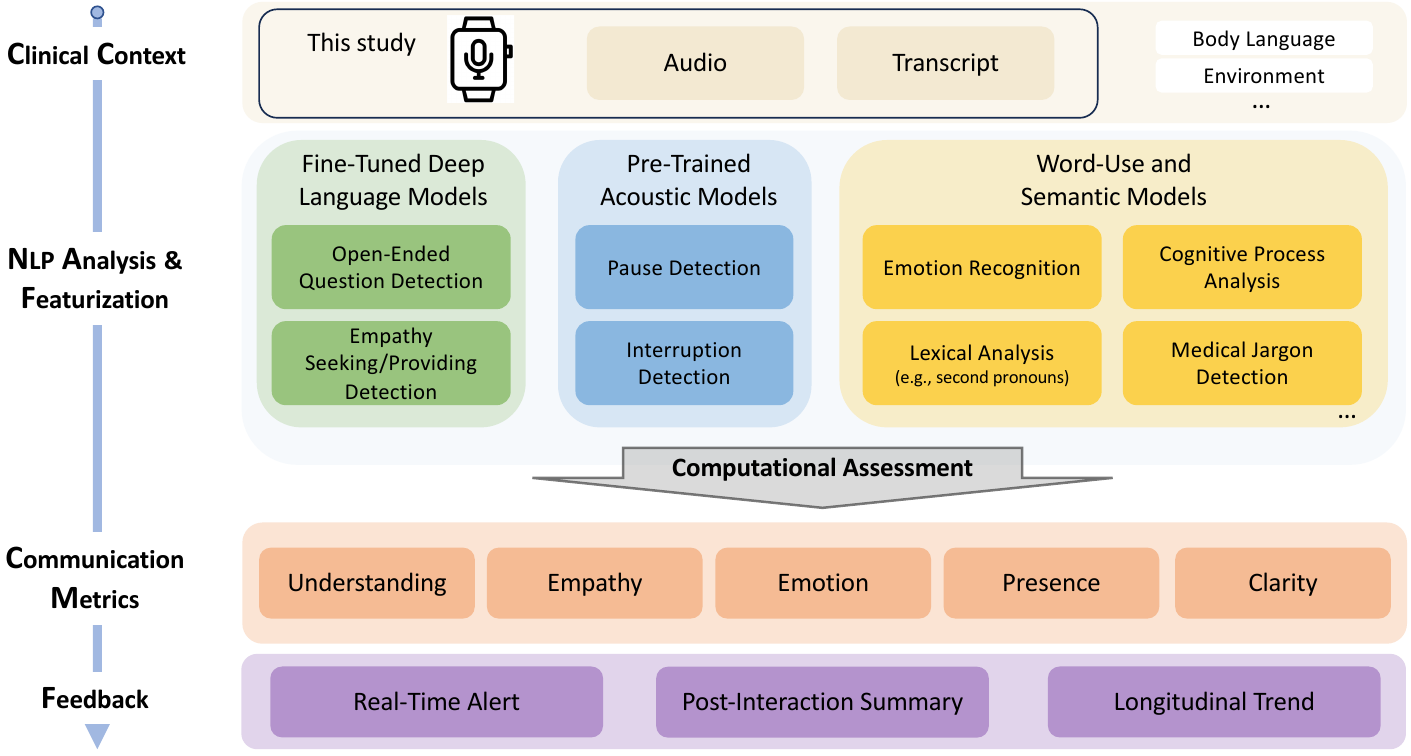}
\caption{A schematic representation of the CommSense Computational Framework, focusing on the fusion of audio and speech-to-text transcription elements. The framework combines clinical and computational aspects, including identification of communication metrics, feedback mechanisms, data sourcing, computational solution deployment, and NLP analysis and featurization. The schematic also suggests future inclusion of additional elements such as body language and environmental factors. This framework is inspired by established computational frameworks in the domains of wearable sensing \cite{mohr2017personal} and medical practices \cite{chen2021cognitive}.}
\label{fig:computational-framework}
\end{figure}

Our computational framework, CommSense, illustrated in Figure \ref{fig:computational-framework}, is designed to extract core communication metrics from wearable sensing data, augmenting the clinical context with communication metrics to provide objective assessment and effective feedback. 

Specifically, the development of CommSense is informed by `best-practice' \textbf{Communication Metrics} (Section \ref{sec:communication-metrics}) established through a comprehensive literature review \cite{lebaron2023commsense}. Starting from the top and moving downwards of Figure \ref{fig:computational-framework}, the framework commences with the \textbf{Clinical Context} (Section \ref{sec:data-source}), leveraging elements feasible to be captured by wearables, such as audio and human speech. This approach is in line with existing research like Wittenberg \textit{et al.} \cite{wittenberg2018communicating} and Ferrell \textit{et al.} \cite{ferrell2020enhancing}, who emphasizes the importance of contextual elements in communication. Next, it employs a variety of \textbf{NLP analysis and featurization} (Section \ref{sec:nlp-techniques}), including deep language models, pre-trained acoustic models, and word-use and semantic models to analyze the audio and transcript data. These techniques facilitate a detailed evaluation of both linguistic and paralinguistic communication cues, providing a comprehensive computational feature set for assessment. Then, following \textbf{Computational Assessment} (Section \ref{sec:rules}) drawn from existing literature and professional guidance, these features are used to determine the qualitative aspects (good or bad) of the five communication metrics per interaction segment. Finally, at the bottom, the computational findings are turned into a diverse range of \textbf{Feedback} (Section \ref{sec:feedback}), providing valuable insights to guide current and future interactions.

\subsection{Establishing Best-Practice ``Communication Metrics''} \label{sec:communication-metrics}

In developing the CommSense framework, our aim was to define communication metrics that not only reflect effective, high-quality provider communication as per medical guidelines but also lend themselves to measurement and extraction through digital technologies and advanced NLP techniques. This identification process, detailed in prior work \cite{lebaron2023commsense}, involved our clinical team members systematically reviewing relevant literature to pinpoint current 'best practices' in palliative care and serious illness communication. This comprehensive review, complemented by inputs from external communication experts, led to the consensus on five pivotal communication metrics.

These metrics, as outlined in Table \ref{tab:summarization-commsense}, include:

\begin{itemize}
    \item Understanding: Efforts focused on exploring and seeking patient understanding \cite{wittenberg2018comfort,wittenberg2018communicating,geerse2019qualitative,ferrell2020enhancing}.
    \item Empathy: The provider's ability to demonstrate empathy in interactions \cite{long2011ten, boyle2017palliative, moore2018communication, jain2020goals, pajka2021feasibility, schwartz2021physician}.
    \item Emotion: Skills in perceiving and reacting to patient emotions \cite{back2007efficacy, kwok2020evaluation, pajka2021feasibility, jain2020goals, back2020patient, wittenberg2018communicating, sommovilla2019discussing}.
    \item Presence: The ability of being present in patient interactions \cite{long2011ten, bernacki2014communication, wittenberg2018communicating, moore2018communication, sommovilla2019discussing, geerse2019qualitative, paladino2020patient}.
    \item Clarity: Clear and understandable communication delivery \cite{baile2005patient, long2011ten, mazor2013patients, gilligan2018patient, wittenberg2018communicating, martin2019communication, paladino2020patient, back2020patient}.
\end{itemize}

These metrics are central to the automation and enhancement of communication in clinical settings \cite{wittenberg2018comfort, wan2016modeling}, utilizing existing methodologies and datasets (e.g., \cite{AmazonQA, hosseini2021takes})
for practical implementation. To establish a technical solution that aligns with these identified metrics, we proposed various strategies and examples for their operationalization, ensuring that the solutions are technically feasible from a computational perspective.

\begin{table}[]
\caption{Summary of the guiding 'Best Practice' principles for quality patient-provider communication in the design of CommSense, outlining metrics with citations, identifiers, operationalization concepts, implementation tools, and rules (adapted from \cite{lebaron2023commsense}). For instance, we evaluate the clinician's ability to explore/seek understanding (tagged as "UNDERSTANDING") by detecting the usage of open-ended questions using an end-to-end model. A segment containing detected open-ended questions is classified as exhibiting "Good Understanding".}
\label{tab:summarization-commsense}
\scriptsize
\begin{tabularx}{1\textwidth}{
  >{\hsize=0.55\hsize}X
  >{\hsize=0.85\hsize}X
  >{\hsize=1\hsize}X
  >{\hsize=1.3\hsize}X
  >{\hsize=1.3\hsize}X
  }
\hline
\textbf{Metric} & \textbf{Grount Truth Tags} & \textbf{Operationalization Concepts} & \textbf{Computational Tools for Implementation} & \textbf{Rules for Computational Implementation} \\ \hline
Understanding \cite{wittenberg2018comfort,wittenberg2018communicating,geerse2019qualitative,ferrell2020enhancing} & 
UNDERSTANDING (good) & 
Employment of open-ended questions/ statements. & 
End-to-end open-ended question detection \cite{wan2016modeling}. \newline \gray{(Training machine learning models on Amazon Question and Answer Dataset \cite{AmazonQA})} & 
{[}Good Understanding{]} \newline Open-ended question statements detected
\\ \hline
Empathy \cite{long2011ten,long2011ten,boyle2017palliative,moore2018communication,jain2020goals,pajka2021feasibility,schwartz2021physician} & 
EMPATHY (good) \newline \gray{Name; Understand; Respect; Support} & 
Utilization of specific phrases/ words & 
End-to-end empathy seeking/ providing detection \cite{hosseini2021takes}. \newline \gray{(Training machine learning models to categorize statements as empathy seeking, empathy providing, or neutral using a public dataset \cite{MahhosEmpathy})} & 
{[}Good Empathy{]} \newline Empathy-providing statements detected 
\\ \hline
Emotion \cite{back2007efficacy,kwok2020evaluation,pajka2021feasibility,jain2020goals,back2020patient,wittenberg2018communicating,sommovilla2019discussing} & 
EMOTION (good) \newline \gray{Respond/Allow};  \newline \newline EMOTION (bad) \newline \gray{Dismiss/ Deflect/ Intellectualize} & 
$\circ$ Avoid deflecting/ dismissing/ changing the subject; \newline $\circ$  Respond to emotion with factual information; \newline $\circ$ Pause for 10 seconds following the delivery of difficult news (allowing time for patient emotional processing) & 
$\circ$ Detection of \newline 1) emotional tone concordance; \newline 2) cognitive response to emotional statements; \newline using Linguistic Inquiry Word Count (LIWC) analysis \cite{tausczik2010psychological}: \newline \gray{emotion level; cognitive process level; emotional tone (positive/ negative);} \newline $\circ$ Silence detection and acoustic analysis for proper pause detection & 
{[}Good Emotion{]} \newline 1. Clinician's emotion direction (positive vs. negative) aligns with the patient's \newline 2. 10-second pause after a statement with negative emotion; \newline {[}Bad Emotion{]} \newline 1. Respond to high emotion level statement with high cognitive process level statement; \newline 2. No pause after a statement with highly negative emotion 
\\ \hline
Presence \cite{long2011ten,bernacki2014communication,wittenberg2018communicating,moore2018communication,sommovilla2019discussing,geerse2019qualitative,paladino2020patient} & 
SILENCE (good) \newline ACTIVE LISTENING (good); \newline \newline SILENCE (bad) \newline INTERRUPTION (bad) & 
$\circ$  Encourage silence / pauses <50\% of the conversation; \newline $\circ$ Active listening [Reflection/ confirmation of understanding/ essential information restatement/ paraphrasing]; \newline $\circ$ Avoid interruptions/ talking over & 
$\circ$ Detection of speech ratio, pauses, and interruptions via acoustic analysis: \newline 
\gray{voice identification; silence detection; overlap detection} \newline $\circ$ Paraphrase identification based on semantic similarity \cite{yin2015convolutional} & 
{[}Good Presence{]} \newline 1. (Overall) Encourage silence; \newline 2. (Overall) Engage in conversation < 50\% of the time; \newline 3. Detected paraphrasing indicated by semantic similarity; \newline {[}Bad Presence{]} \newline Interruptions detected
\\ \hline
Clarity \cite{baile2005patient,long2011ten,mazor2013patients,gilligan2018patient,wittenberg2018communicating,martin2019communication,paladino2020patient,back2020patient} & 
LANGUAGE (good) \newline \gray{Explained/ Clear}; \newline \newline LANGUAGE (bad) \newline \gray{Unexplained/ Unclear} & 
$\circ$ Utilize plain, direct language [active voice, use second person, sentences <15 words]; \newline $\circ$  Limit medical jargon; \newline $\circ$  Contextualize/ indicate next steps & 
$\circ$ Detection of plain language using LIWC analysis: \newline \gray{word and sentence count; words per sentence; cognitive process level; usage of first and second pronouns} \newline $\circ$ Detection of medical jargon based on medical dictionary \cite{HarvardMedicalDictionary} & 
{[}Good Clarity{]} \newline 1. (Overall) Use more second pronouns and less first pronouns than usual; \newline 2. Using medical jargon with explanation provided; \newline {[}Bad Clarity{]} \newline 1. Detected overuse of medical jargon; \newline 2. Long sentences (words per sentence > 15)
\\ \hline
\end{tabularx}
\end{table}

\subsection{Sensing ``Clinical Context'' Using Smartwatches} \label{sec:data-source}

The input layer of the framework (see Figure \ref{tab:summarization-commsense}) is the audio data source, which uses smartwatch \textit{audio and speech-to-text transcript} streams to extract communication metrics. This layer is foundational for capturing real-time interactions and providing data essential for evaluating communication effectiveness in clinical contexts \cite{wittenberg2018comfort,geerse2019qualitative}. During a patient-provider interaction, the smartwatch is   programmed to collect and analyze audio data. Data collection is activated by the provider clicking on the screen to start or stop the recording. Note that this can be automated in future work by automatically detecting relevant context by combining the audio data with other contextual information such as location, time, and calendar data to automatically detect relevant patient-provider interaction episodes (e.g., audio interactions happening in the hospital and overlapping with patient visit calendar event).

\subsection{Enabling ``NLP Analyses \& Featurization''} \label{sec:nlp-techniques}

To operationalize the metrics and process the audio and transcript data, we employed a variety of state-of-the-art linguistic analysis and NLP techniques. These techniques serve different functions within the framework, from identifying open-ended questions and empathy-providing statements to tracking acoustic elements such as pauses and interruptions. A subset of the techniques pay particular attention to word usage and semantic analysis, enabling the system to comprehend emotional context, cognitive processes, and decipher medical terminology. The integration of these techniques provides an innovative approach to quantifying complex communication skills, offering a new lens through which we can understand, measure, and ultimately enhance healthcare provider-patient interactions. Through this systematic and data-driven approach, we aim to redefine how communication quality is assessed and improved, by showcasing the linguistic and acoustic aspects that contribute to effective communication.

\subsubsection{Fine-Tuned Deep Language Models}

The incorporation of advanced deep learning in our framework empowers us to automatically evaluate several pivotal features in an end-to-end manner, including open-ended questions and empathy seeking or providing interactions. We opted to use a BERT-based fine-tuned model \cite{devlin2018bert}, a powerful tool that leverages the strength of a pre-trained deep language model, trained on a vast corpus of text (like Wikipedia and BooksCorpus) and already familiar with the intricate facets of language, including syntax (sentence structure and grammatical relationships), semantics (the meaning of words and sentences), and context-based nuances (the different meanings words can have depending on the context). By adding a classification layer on top, we fine-tune the model to our specific task, thereby allowing for accurate, nuanced interpretations of various communication contexts. This scalable and increasingly precise method is demonstrated through the detection of open-ended questions and empathy-seeking/providing interactions:

\begin{itemize}
\item \textbf{Open-ended questions}: To identify the communication metric of \textit{seeking understanding}, we focused on the detection of open-ended questions which are used to elicit a range of detailed responses, as opposed to simple affirmations or negations. These questions facilitate deeper exploration of the topic versus yes-or-no answers, which are identifiable through recent advances in text classification \cite{wan2016modeling, AmazonQA}. For this purpose, we fine-tuned the pre-trained BERT model on the Amazon Question/Answer dataset\footnote{Amazon Question/Answer data: \url{https://cseweb.ucsd.edu/~jmcauley/datasets/amazon/qa/}}, which comprises 80,496 healthcare-related questions. These questions are categorized by ``question type'' labels, distinguishing between open-ended and polar (yes/no) question.

\item \textbf{Empathy seeking/providing}: To evaluate the metric of \textit{conveying empathy}, the BERT model was fine-tuned to pinpoint instances of both ``empathy seeking'' (from the patient's side) and "empathy providing" (from the provider's side). Following the recent advances \cite{hosseini2021takes,MahhosEmpathy}, we utilized the Empathy dataset\footnote{Empathy dataset: \url{https://github.com/Mahhos/Empathy}}, which contains labeled sentences for ``seeking empathy'', ``providing empathy'', and ``neutral''. Consequently, our refined model proficiently predicts empathy-seeking and providing interactions. 
\end{itemize}

These fine-tuned, BERT-based language models exemplify a data-driven strategy for identifying and enhancing essential communication metrics. While traditional methods of "empathy" operationalization, such as detecting certain phrases or words (as listed in Table \ref{tab:summarization-commsense}), can be challenging due to the inherent abstractness of empathetic language, the fine-tuned models offer a compelling solution for identifying these nuanced communication metrics in text.

\subsubsection{Acoustic Analysis}

The acoustic properties of the recorded audio data also offer essential insights for evaluating the communication metrics.

\begin{itemize}
    \item \textbf{Pauses}: Pauses in patient-provider conversations can indicate moments of \textit{allowing emotion}; for example, a 10-second pause following the delivery of difficult news is considered effective \cite{back2007efficacy}. Our methodology for detecting pauses combines audio loudness analysis and silence detection. We compute the volume for each audio frame, and define a silence threshold using an adaptive percentile-based approach (in this case, we set the 20th percentile of the audio frame volumes as the threshold to detect silence) for each data sample. Silent intervals are then mapped onto the conversation timeline. By comparing silence intervals with conversation segment start times from the transcript, we can accurately identify and measure pauses and the portions they belong to.
    \item \textbf{Interruptions}: Interruptions can indicate to an undesired way to \textit{be present} and can prevent patient-clinician connection. To accurately detect speech overlap, which we define as interruptions, we employ a pretrained deep learning model\footnote{Pre-trained overlapped speech detection: \url{https://huggingface.co/pyannote/overlapped-speech-detection}} \cite{Bredin2021}. The pipeline is applied to the audio data to identify instances of overlapping human voice. These instances are converted into time intervals and mapped onto the conversation timeline. Using timestamps from the conversation's transcript, we compare the identified overlaps with the start times of conversation segments. An overlap detected within a specific segment suggests an interruption occurred during that period.
\end{itemize}

\subsubsection{Word-use and semantic analysis}

We also incorporate word-use and semantic analysis to further enrich the understanding of the communication metrics. We used the Linguistic Inquiry and Word Count (LIWC)\footnote{LIWC-2022: \url{https://www.liwc.app/}} analytics tool to extract linguistic features of each portion of the conversation, with additional methods such as medical terminology detection as a supplement. These techniques are vital for a comprehensive assessment of communication, helping to identify key aspects such as emotion recognition, cognitive process, lexical analysis, speech dominance, and medical jargon use, which are integral to effective communication \cite{tausczik2010psychological, HarvardMedicalDictionary, yin2015convolutional}.

\begin{itemize}
    \item \textbf{Emotion recognition}: One of the central features we focus on is the detection of the polarity (positive or negative) and the magnitude of the patient's and the provider's \textit{emotion}. To achieve this, we leverage LIWC's sentiment analysis capability. The tool provides emotion-related linguistic feature counts that give insights into the emotional tone of a conversation, for both the provider and the patient. By analyzing these emotion-related features, including positive emotion, negative emotion, anger, anxiety, and sadness, we can gauge the emotional dynamics and coherence within which the conversation is taking place. This method evaluates the ``Emotion'' communication metric, helping to understand if the healthcare provider is appropriately responding to the patient's emotional cues \cite{tausczik2010psychological}. 
    \item \textbf{Cognitive process}: In communication, particularly in healthcare, it's crucial to \textit{respond to emotions} with empathy and understanding, rather than solely relying on cognitive facts or figures. A patient expressing emotions is typically seeking emotional support, rather than raw data or facts \cite{jain2020goals}. If a provider responds merely with facts or figures, it may come across as dismissive or indifferent to the patient's emotional state, leading the patient to feel unheard or misunderstood. To assess the cognitive processes involved in the provider's communication, we use LIWC tool to evaluate the ``Emotion'' metric by detecting unsatisfied cases in which the provider deliver facts/statements with complex cognitive process, thereby determining whether the provider is adequately taking care the patient's emotion.
    \item \textbf{Lexical analysis}: For the conversation with \textit{high clarity}, the use of second-person pronouns and a lower average word count per sentence (ideally, less than 15 \cite{wittenberg2018communicating}) is encouraged. This can make the provider's communication more personal, direct, and easy to understand for the patient. We operationalize this by implementing lexical analysis techniques on the conversation transcript. Specifically, we count the occurrences of second-person pronouns, like `you' and `your', in the provider's speech as a proportion of total words used. Additionally, we calculate the number of words used per sentence by the provider. These metrics are then compared to our predefined thresholds to assess the provider's performance in terms of ``Clarity'', ensuring the use of language that is accessible and understandable to the patient \cite{back2020patient}.
    \item \textbf{Speech dominance}: The metric ``Presence'' emphasizes the importance of the healthcare provider \textit{actively listening} to the patient and not dominating the conversation. To quantify this, we measure the percentage of the conversation covered by the healthcare provider's own speech. This is achieved through speaker identification techniques, which distinguish between the voices of different speakers in the conversation. If the provider's speaking time exceeds half of the total conversation duration, it is flagged as excessive, indicating the provider may not be adequately listening to the patient's concerns. This analysis helps in ensuring balanced and patient-centric communication \cite{long2011ten, bernacki2014communication}.
    \item \textbf{Medical jargon}: We also identify the use of medical jargon in the provider's speech. \textit{Excessive and unclear usage of medical jargon} can impact the `Clarity' metric, as it may hinder the patient's understanding of the conversation \cite{back2020patient}. To capture potential medical jargon, we utilize the Harvard Medical Dictionary\footnote{Harvard Medical Dictionary: \url{https://www.health.harvard.edu/a-through-c}} as a standard reference. Note that we manually filtered out commonly known terms (such as cation, radiation, vein, and opioid) from the dictionary to avoid false positives in identifying challenging jargon.
\end{itemize}

\subsection{Computational Assessment of Provider Communication} \label{sec:rules}

As shown in Table \ref{tab:summarization-commsense}'s computational tools for automation column, each of these techniques contributes to a comprehensive analysis of the recorded audio data, providing a multifaceted evaluation of the communication metrics. In Table \ref{tab:summarization-commsense}'s Identification Rule(s) column, we detail the computational assessment we propose for each communication metric. The detailed assessments are listed in Table \ref{tab:rules}. 

\begin{table}[t!]
  \caption{Computational assessment rules linking clinical tags and corresponding computational features.}
  \centering 
  \scalebox{1}{
  {\scriptsize

  \begin{tabularx}{\textwidth}{|>{\hsize=.25\hsize}X|>{\hsize=.6\hsize}X|>{\hsize=1.15\hsize}X|}
  \hline
  \rowcolor{gray!30}
\textbf{Metric} & \textbf{Clinical Ground Truth Tags} & \textbf{Computational Features}\\
\hline
\rowcolor{cyan!5}
\textbf{Understanding} & Understanding (good) & Open-ended questions (number of open-ended questions)\\
\hline
\rowcolor{green!5}
\textbf{Empathy} & Empathy (good) - Name/ Understand/ Respect/ Support/ Explore & Providing empathy (number of sentences providing empathy)\\
\hline
\rowcolor{orange!5}
\textbf{Emotion} & Emotion (good) - respond & Emotion/Tone Attitude (value of the positive/negative degree)\\
\cline{2-3}
\rowcolor{orange!5}
& Emotion (good) - allow & Length of Pause (seconds)\\
\cline{2-3}
\rowcolor{orange!5}
& Emotion (bad) - dismiss/ deflect/ intellectualize & Emotion/Tone Negative (value of the negative degree), Cognitive Process (value of the degree of the cognitive process complexity)\\
\hline
\rowcolor{blue!5}
\textbf{Presence} & Silence (good)- good pause & Length of Pause (detected pause from audio)\\
\rowcolor{blue!5}
& Silence (bad)- bad pause & \\
\cline{2-3}
\rowcolor{blue!5}
& Interruption (bad) - interruption & Overlap of Voice (if disrupted or not, yes- 1 no-0)\\
\cline{2-3}
\rowcolor{blue!5}
& Active Listending (good) - active listening & Detected Rephrasing Sentence (Identifies if a sentence in the current section, e.g., provider's speech, rephrases a sentence from the last patient's section)\\
\hline
\rowcolor{red!5}
\textbf{Clarity} & Clarity (good) - explained/clear & Word per Sentence, Big Words, Word per paragraph\\
\cline{2-3}
\rowcolor{red!5}
& Clarity (bad) - unexplained/unclear & Medical Jargon Count and Details\\
\cline{2-3}
\rowcolor{red!5}
& Clarity (good) - active voice & Authenticity/Honesty (Level of the authenticity)\\
\rowcolor{red!5}
& & Positive Tone/Emotion (Value of the positive degree)\\
\cline{2-3}
\rowcolor{red!5}
& Clarity (good) - use second person & I-Statement (Number of the first person words, e.g., I, me, my)\\
\rowcolor{red!5}
& & You-Statement (Number of the second person words, e.g., you, your)\\
\cline{2-3}
\rowcolor{red!5}
& Clarity (good) - sentences < 15 words & Word per Sentence Count (Number of the words)\\
\hline

\end{tabularx}}}
  \label{tab:rules}
\end{table}

Note, the specific parameters for operational rules to determine ``good'' or ``bad'' are typically set according to established literature, in compliance with standards, or by the researchers' manual search and adjustments. For example, literature suggests a 10-second pause following the delivery of difficult news is effective \cite{back2007efficacy}, and that an optimal average sentence length in clinical communication is fewer than 15 words \cite{wittenberg2018communicating}, a standard also recommended in clinical writing practices \cite{aldridge2004writing}. Furthermore, for metrics such as negative emotion and tone, identified using the LIWC tool, we employ standard thresholds indicative of formal language averages provided by the tool. Some thresholds are also tailored the patient-provider communication context versus formal language averages.

\subsection{``Feedback'' To Providers} \label{sec:feedback}

\begin{figure*}[t]
\centering
\subfigure[Main Interface]{%
  \includegraphics[width=0.22\columnwidth]{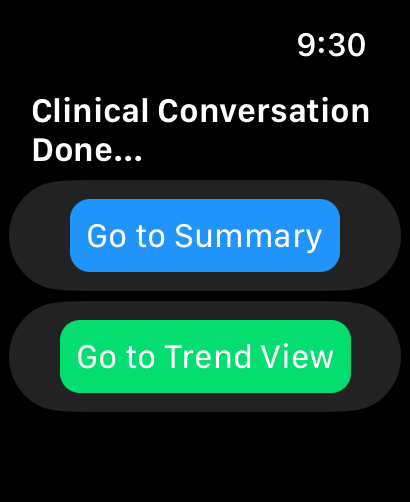}%
  \label{fig:}%
}
\quad
\subfigure[Real-Time Alert]{%
  \includegraphics[width=0.22\columnwidth]{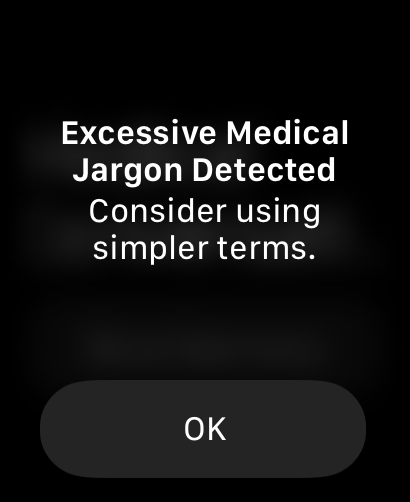}%
  \label{fig:}%
}
\quad
\subfigure[Post-Interaction Summary]{%
  \includegraphics[width=0.22\columnwidth]{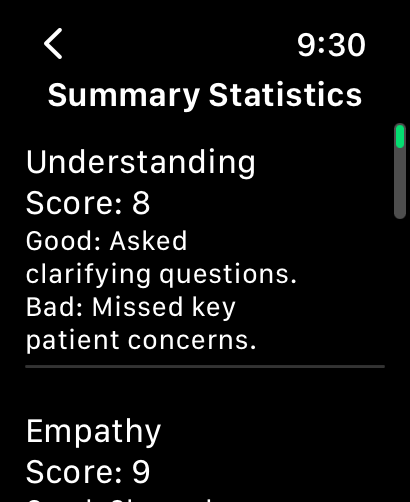}%
  \label{fig:}%
}
\quad
\subfigure[Longitudinal Trend]{%
  \includegraphics[width=0.22\columnwidth]{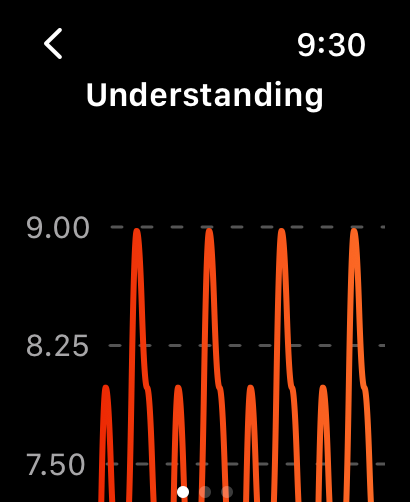}%
  \label{fig:}%
}
\caption{Feedback interface prototypes.}
\label{fig:analysis-1-1}
\end{figure*}

Feedback in the context of communication training and practice for healthcare providers is essential to facilitating improvements and reinforcing the effective use of ``best-practice'' communication metrics. This feedback can be delivered through various user interfaces (UIs) designed for different purposes, including real-time feedback during the interaction, post-interaction summary statistics, and longitudinal data showing trends in performance over time. Here we propose designs for these three types of feedback interfaces that we plan to implement in future work:

\subsubsection{Real-Time Feedback}

Real-time feedback can be delivered to the provider using smartwatch pop-ups, which can serve as an immediate prompt to adjust their communication style as suggested. The design for this interface should be non-disruptive, to avoid detracting from the flow of the patient-provider conversation. For instance, when excessive medical jargon is being used, a subtle vibration alert on the smartwatch can signal the provider to modify their speech to enhance clarity or explanation. Moreover, a minimalistic visual cue (e.g., a pop-up notification) on the smartwatch screen can further specify the kind of communication metric the provider needs to adjust.

\subsubsection{Post-Interaction Summary Statistics}

Post-interaction feedback is vital to allow healthcare providers to reflect on their communication performance. This feedback can be delivered through an application interface, presenting a summary of communication metrics in a simple format. The design of this interface can incorporate a composite scorecard, highlighting scores for each communication metric based on the CommSense analysis. The score can be presented as a numerical value, along with a color-coded or star-rated system for quick visual interpretation. Additionally, to provide context, this dashboard can include specific instances of conversations that contributed to the scoring for each metric.

\subsubsection{Longitudinal Feedback}

Longitudinal feedback can offer healthcare providers insights into their communication performance trends over time. This can be achieved by displaying progress graphs or charts that show how each metric has improved or declined over a specified period (e.g., in the past month). For example, to show improvement, a line graph can display the trend of `Understanding' scores over the past month. The graph should clearly denote key points where significant changes occurred, perhaps associated with specific training sessions or notable interactions.

Moreover, longitudinal feedback can also serve as a valuable tool for instructors responsible for overseeing trainee development. By making use of the data gathered and analyzed by CommSense, instructors can gain a granular understanding of the trainees' communication skills progress over time. It can highlight areas of strength and opportunities for growth, allowing instructors to tailor their coaching and training modules according to individual needs.

\section{Pilot Study to Establish Technical Proof of Concept and Feasibility}

To field-test the proposed CommSense framework and gather valuable insights, we undertook a pilot feasibility study with a sample of nursing and medical students (N=35) and practicing palliative care providers (N=5) in a simulated clinical setting. We aimed to confirm end-user acceptability and to ascertain technical efficacy in capturing nuanced clinical communication metrics. In this paper, we focus on our work to validate the technical feasibility of CommSense, as guided by the following research questions: 
\begin{itemize}
    \item[\textbf{RQ1:}] Is the idea of CommSense feasible and acceptable for evaluating palliative care communication?
    \item[\textbf{RQ2:}] How effectively does the proposed CommSense framework capture the communication metrics, and what areas require improvement in future iterations?
\end{itemize}
\subsection{Participants}

This pilot study recruited N=40 participants from an academic medical center using a strategic snowball sampling approach. Participants included pre-licensure nursing and medical students and licensed practicing clinicians to test the CommSense framework during sessions lasting approximately 20–30 minutes each. Participants were offered a \$25 gift card as compensation.

This group comprised five practicing clinicians (12.5\%), nineteen nursing students (47.5\%), and sixteen medical students (40\%). The majority were in the age group 18-24 (N=29, 72.5\%),  identified as female (N=32, 80\%). In terms of self-reported racial identity, 60\% were white, 5\% were black/African American, 37.5\% were Asian, and 2.5\% selected the option of ``other''. Participants could identify multiple racial categories. In terms of healthcare roles and disciplines, participants included both physicians and nurses. Specifically, we had physicians (including medical students) making up 19 participants (47.5\%) and nurses (including nursing students and nurse practitioners) forming a group of 21 participants (52.5\%).

\subsection{Pilot Study Settings}

For data collection, we employed the Huawei Smartwatch 2 and a securely programmed mobile crowdsensing platform based on SWear \cite{boukhechba2020swear}, to gather audio signals\footnote{We also collected physiological data such as accelerometer data for future non-verbal analysis.} in the simulated clinical environment. SWear, chosen for its robust data encryption and secure data transfer capabilities, ensures the protection of sensitive audio data.

The study was conducted with the approval of the Institutional Review Board (IRB) of a U.S. public university. All participants provided written informed consent prior to data collection. This process provided participants with full information about the study’s objectives, procedures, potential risks and benefits, confidentiality protocols, and the intended use of the data. All collected data were anonymized and securely stored on a high-security IRB-approved data server, with access strictly limited to authorized researchers. These measures collectively ensured the highest standards of privacy and data security were maintained throughout the study.

\subsection{Script and Clinical Tag Development} \label{sec:scripts}

\begin{figure*}[t]
\centering
\subfigure[A fragment of a ``Good'' script between the patient and the physician.]{%
  \includegraphics[width=0.9\columnwidth]{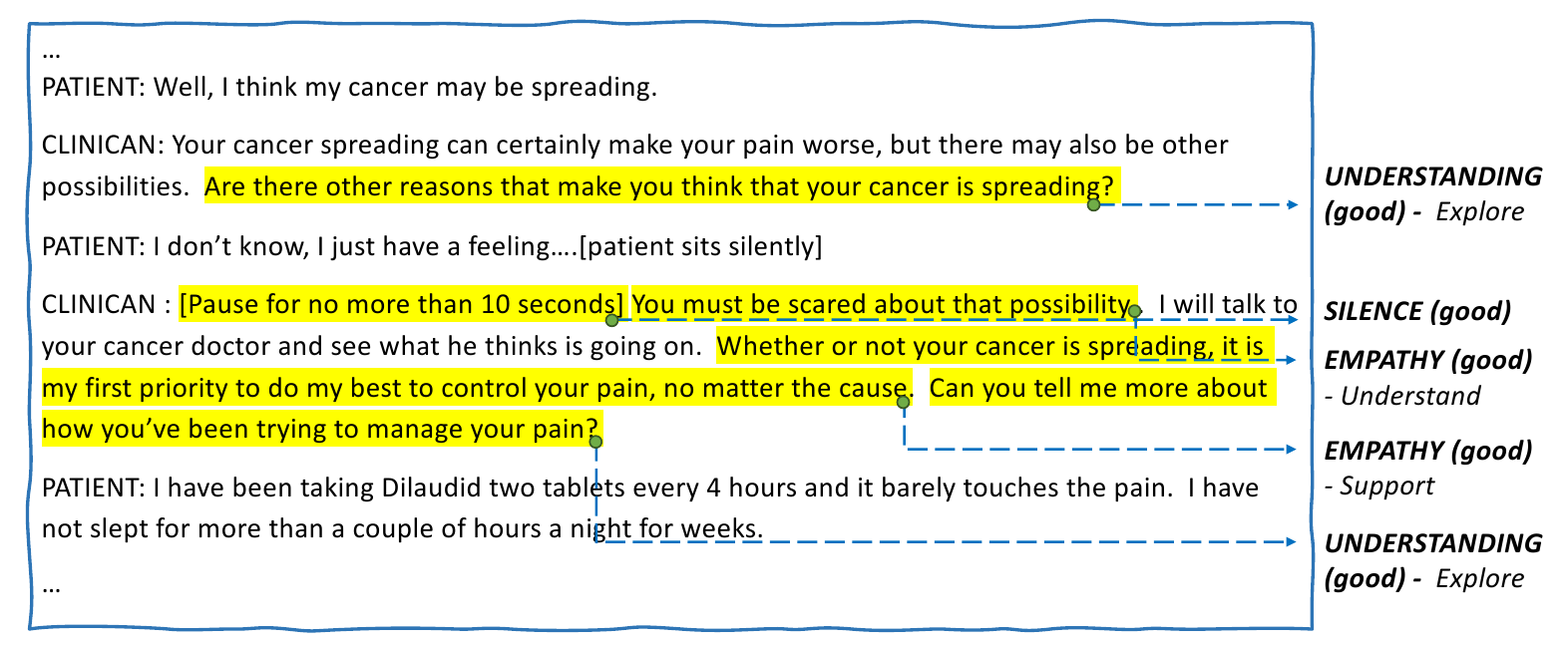}%
  \label{fig:}%
}
\subfigure[A fragment of a ``Bad'' script between the patient and the nurse.]{%
  \includegraphics[width=0.9\columnwidth]{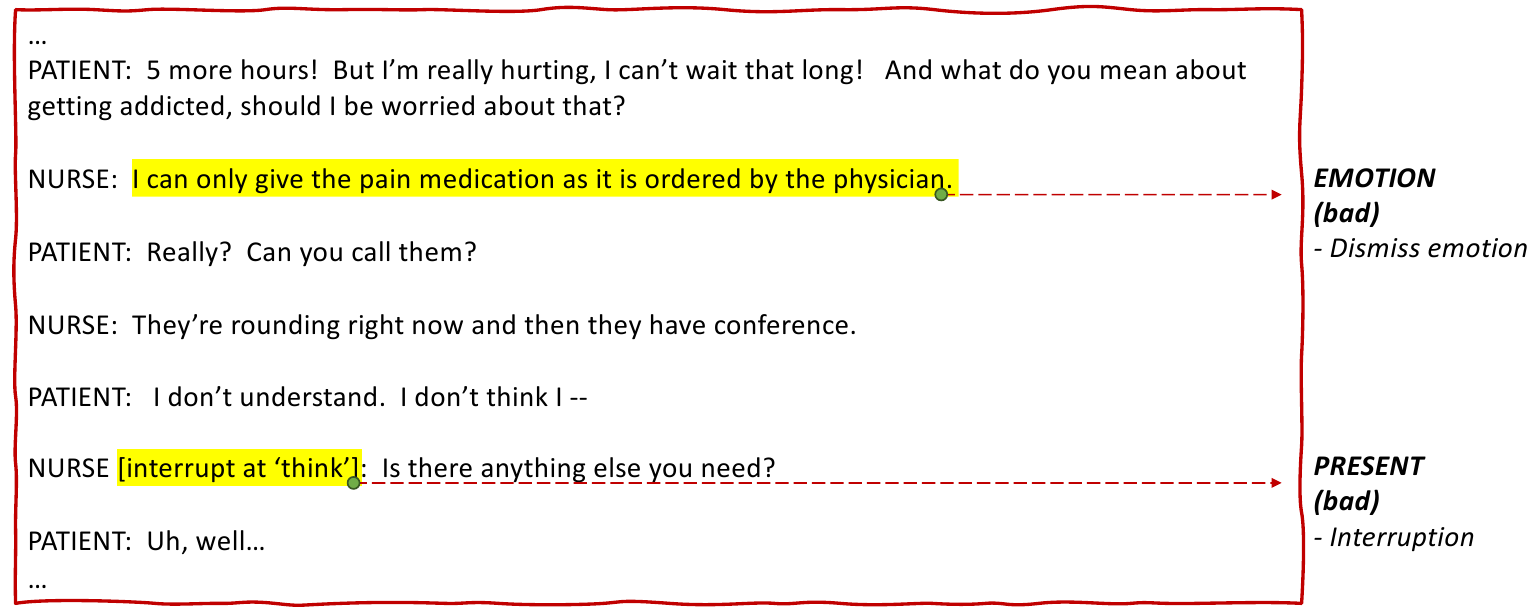}%
  \label{fig:}%
}
\caption{Example fragments with the `Good'' and ``Bad'' tags in scripted generated.}
\label{fig:example-scripts}
\end{figure*}

To assess the framework's capability in identifying the defined metrics, a collaborative effort between our clinical and engineering teams led to the creation of Ground Truth conversation scripts embedded with our predefined communication metrics (refer to Section \ref{sec:communication-metrics}), for pilot testing. These scripts were developed with an emphasis on advanced cancer pain management and the discussion of prognosis and care goals.

Eight scripts were created in total. Specifically, two distinct types of scripts were created, one tailored for physicians and the other for nurses. Each set encompassed two ``good'' conversations with desired/positive communication metrics like empathy, and two ``bad'' conversations with negative communication metrics such as inappropriate use of medical jargon or interruptions. The eight conversation scripts were meticulously crafted by clinical research team members, and then independently tagged with the appropriate metrics. To address the inclusion of negative communication, sections of the dialogue that embodied the key metrics were marked as either ``good'' or ``bad''. Segments that did not distinctly display any of our predefined metrics were left untagged. For the five communication metrics we used, sections in the ``good'' scripts were tagged as either ``Good'' or left untagged, while sections in the ``bad'' scripts were tagged as ``Bad'' or also left untagged. Any disagreements were addressed through group discussions until consensus was reached.

\subsection{Data Collection and Processing}

In the simulated clinical setting, our study participants assumed the roles of healthcare providers and were tasked with reading conversation scripts that matched their discipline (e.g. nurses and nursing students read the nurse scripts; physicians and medical students read the physician scripts). Depending on the participants' time availability, they read aloud one or two ``good'' scripts, typically lasting 6 minutes, and one or two ``bad'' scripts, which took about 2-3 minutes. In these scripted interactions, members of our research team — consisting of clinical faculty and students — assumed the roles of patients. Both the participant (acting as the healthcare provider) and the researcher (as the patient) wore a programmed smartwatch. While our experiment design required only the provider's watch to record data, we included this redundancy as a safety net to prevent potential data loss.

In the post-session data collection, we asked participants to complete surveys related to their professional roles, their willingness to use CommSense, and their preferences regarding feedback content and format (see Section \ref{sec:survey_results} for survey details and results).

For the purpose of pilot-testing, we applied digital translation tools to convert the gathered audio data into transcripts, facilitating further analysis. Specifically, we applied Otter.ai\footnote{Otter.ai \url{https://otter.ai/}} to automatically categorize and label each section as either "provider" or "patient", also providing corresponding timestamps. These timestamps facilitated the segmentation of audio recordings into provider and patient sections.

\section{Results}

In this section, we present the comprehensive results from our examination of the technical feasibility and capability of CommSense. Specifically, we assess the methodologies proposed for each communication metric.

\subsection{Evaluation Strategy}


In our evaluation of CommSense, the first step involves processing the audio and transcript data from each recorded conversation using CommSense analysis and featurization methods. This allows us to extract specific features for each part of the conversation, which we refer to as a `segment'. For the purposes of our study, we define a `segment' as a part of the conversation where one person speaks, irrespective of the number of sentences involved. Following this, we apply computational implementation rules to convert these features into corresponding `good', `bad', or `None' labels of each segment, as a set of ``model outputs'', for each communication metric associated with each conversation segment. Correspondingly, we also organize the ``clinical tags'' manually labelled at segment level to compare with. In the final stage, we compare the automatically generated labels by CommSense (model outputs) with the manually labelled tags (clinical tags) segment by segment.

In assessing the performance of CommSense with the model outputs and clinical tags, we employed \textit{balanced} accuracy, precision, and recall as our key evaluation metrics.

Balanced accuracy, an extension of the traditional accuracy metric, provides a comprehensive perspective on our system's overall performance. It accounts for both true positive and true negative rates, delivering a more balanced performance measurement even when the classes are imbalanced. This metric reveals the proportion of total correct predictions, ensuring both types of correct predictions are given equal importance.

\begin{equation}
\small
\text{Balanced Accuracy} = \frac{1}{2} \left( \frac{\text{True Positives}}{\text{True Positives} + \text{False Negatives}} + \frac{\text{True Negatives}}{\text{True Negatives} + \text{False Positives}} \right)
\end{equation}

Alongside balanced accuracy, we employed two additional key metrics: precision and recall. Precision is used to evaluate how many of the predicted positive instances are actually positive. This metric is particularly important as it helps us to understand the extent to which our model generates false positives. On the other hand, recall assesses the system's capability to correctly identify positive instances among all actual positive instances. In essence, it gives us an understanding of how well the system is able to detect rightly from each set of the tags.

\begin{equation}
\small
\text{Recall} = \frac{\text{True Positives}}{\text{True Positives} + \text{False Negatives}}, \quad \quad \text{Precision} = \frac{\text{True Positives}}{\text{True Positives} + \text{False Positives}}
\end{equation}

\subsection{Evaluation Results}

\begin{table}[t!]
\centering
\caption{Evaluation results of the CommSense model output using Ground Truth tags.}
\small
\label{tab:evaluation}
\arrayrulecolor{black}
\renewcommand{\arraystretch}{1.1}
\begin{tabular}{>{\bfseries}lcccccc}
\toprule
 & \multicolumn{3}{c}{\textbf{Good Scripts}} & \multicolumn{3}{c}{\textbf{Bad Scripts}} \\
\cmidrule(lr){2-4} \cmidrule(l){5-7}
 & Accuracy & Precision & Recall & Accuracy & Precision & Recall \\
\midrule
Understanding & 0.755 & 0.733 & 0.792 & ---\footnotemark & --- & --- \\
Empathy & 0.660 & 0.679 & 0.663 & --- & --- & --- \\
Emotion & 0.861 & 0.733 & 0.846 & 0.729 & 0.721 & 0.678 \\
Presence & 0.862 & 0.723 & 0.703 & 0.797 & 0.835 & 0.558 \\
Clarity & 0.776 & 0.614 & 0.567 & 0.795 & 0.664 & 0.813 \\
\bottomrule
\end{tabular}
\end{table}
\footnotetext{There is no ``bad'' Understanding and Empathy groundtruth based on our design, so that there is no evaluation result about the two.}

Our evaluation of the CommSense system is outlined in Table \ref{tab:evaluation}, focusing on accuracy, precision, and recall scores of detecting ``good'' tags (e.g., empathy good) across the communication metrics from the ``Good Scripts'' and detecting ``bad'' tags (e.g., interruption bad) from the ``Bad Scripts''.

Regarding ``Good Scripts'', the metrics Understanding, Empathy, Presence, Emotion, and Clarity demonstrated varying levels of feasibility. Understanding performed impressively with an accuracy of 75.5\%, implying effective recognition and categorization in conversations. Despite this, a slightly lower precision score for Understanding (73.3\%) signals the potential for refining the system to minimize false positives. On the other hand, Empathy demonstrated lower accuracy (66.0\%), precision (67.9\%), and recall (66.3\%), signaling potential for enhancement. This aligns with the findings of existing work \cite{davoudi2022using} who highlighted the challenges of quantifying complex communication aspects such as empathy using NLP technologies. Another potential factor to this challenge is the nascent state of empathy detection in NLP \cite{hosseini2021takes}, coupled with a lack of dedicated datasets, which complicates the training of effective models. This leads to poorer performance in empathy detection compared to understanding, despite both being based on models developed by our team. Additionally, Presence and Emotion emerged as the top-performing metrics in the ``Good Script'', scoring an accuracy of 86.2\% and 86.1\%, respectively. Yet, their precision and recall numbers, at 72.3\% \& 70.3\% and 73.3\% \& 84.6\%, respectively, indicate areas for improving the detection of ``good'' presence. Furthermore, Clarity, with the lowest precision (61.4\%) and recall (56.7\%), signifies scope for reducing false positives and improving the system's ability to correctly identify `good' Clarity where it truly exists.

For the ``Bad Scripts'', the metrics Presence, Emotion, and Clarity were evaluated. Understanding the negative communication metrics is essential as they can hinder patients' trust and lead to complications \cite{tiwary2019poor}. Note, Understanding and Empathy were not included in the ``Bad Script'' evaluation as our ground truth design did not incorporate ``bad'' instances for these metrics. However, the lower recall for Presence (55.8\%) suggests that enhancements are needed in identifying actual negative instances. Interestingly, Clarity demonstrated a robust recall rate (81.3\%) but lower precision (66.4\%), pointing to an area for improvement in reducing false positives for unclear communication.

In analyzing the performance discrepancies between `good' and `bad' scripts, it is important to recognize that these scripts encompass two distinct algorithmic targets. For instance, evaluating `good' Presence is contingent on the algorithm's ability to detect adequate pauses and active listening, while `bad' Presence depends on identifying instances of insufficient pauses and interruptions. These fundamentally different objectives naturally lead to variations in the algorithms' performance. Additional complexities arise due to the practical difficulties in accurately identifying `bad' tags and individual variation and biases in script enactment among participants. These considerations underscore the nuances of applying our algorithm and highlight the opportunities for ongoing refinements to accurately discern positive and negative communication aspects and the critical importance of contextualizing CommSense output results.

Moreover, the quality of participant enactment significantly influences the evaluation outcomes. This phenomenon was particularly pronounced in the `Bad Scripts', as evidenced in the evaluation of Presence. Several `bad' tags were designated based on the presence of interruptions, technically characterized by overlapping speech after the patient's statements. However, as we observed, more than half of the participants were unable to accurately enact these `interruptions' during the script reading, leading to a less prominent presence of `interruptions' in the collected data. This limitation in enactment meant that, despite our algorithm's correct identifications, its performance was deemed suboptimal due to the flawed `ground truth' set on incorrect expectations. This was particularly highlighted in the recall for `Bad Scripts' (0.558). Conversely, participants likely portrayed `good' scenarios more effectively, thereby providing a more accurate ground truth for these scripts. This difference in enactment quality significantly contributed to the discrepancy observed in our evaluation outcomes for `good' and `bad' scripts.

Our methodological framework introduces a pioneering analysis of five systematically identified metrics and operationalization rules, paving the way for future research in the area of patient-provider communication. This work has the potential to establish our consensus performance metrics as `standard' metrics and to establish performance benchmarks for further study in the field \cite{davoudi2022using,lebaron2022exploring,tabaie2023natural}, while also highlighting areas for improvement.

\subsection{Survey Results} \label{sec:survey_results}

Post-session survey results show that most participants believe CommSense can enhance patient-clinician communication awareness, improve healthcare quality, and foster compassionate dialogues between providers and patients or their families. A detailed account of the post-study survey findings is provided in \cite{lebaron2023commsense}.

Specifically, the survey revealed that a significant majority, 87.5\% of participants (N=35) are comfortable with the idea of CommSense regarding questions ``I am comfortable with the idea of CommSense recording my clinical conversations.'' and ``I am comfortable with the idea of CommSense assessing my communication skills.'', respectively. However, privacy concerns were noted by 30\% (N=12) when asked if ``CommSense makes me concerned about privacy/confidentiality'', underscoring the necessity for enhanced privacy technology and clear user consent protocols. Despite these concerns, 92.5\% of participants (N=37) expressed a willingness to use CommSense by agreeing with ``I would be willing to use CommSense in the future''. Furthermore, regarding preferred format of ``Feedback from CommSense'', a notable 97.5\% of participants (N=39) agreed that ``could increase awareness of the importance of patient–clinician communication'', 97.5\% of participants (N=39) agreed it ``could help healthcare organizations deliver higher-quality healthcare'', and 95\% of participants (N=38) agreed it ``could help clinicians communicate more compassionately with patients and their family members''. In terms of feedback preferences, post-interaction summary statistics were most favored (95\%, N=38), followed by longitudinal data showing trends (77.5\%, N=31). In contrast, real-time feedback during interactions was least preferred, with only 22.5\% (N=9) favoring it.

\section{Discussion}

In this section, we provide a comprehensive summary of the implications of CommSense, including its technical and clinical implications, as well as possibilities for future enhancements and refinements.

\subsection{Technical Implications}

The development and pilot testing of the CommSense framework have brought to the forefront several important technical implications. This work, as a pioneering attempt at leveraging wearable-enabled computational sensing for assessing patient-provider communications, offers valuable insights into the realm of digital health technology.

Firstly, the results underscore the feasibility of utilizing ubiquitous sensing and natural language processing techniques to extract core communication metrics from patient-clinician interactions. This study confirms that technologies such as automated speech recognition, emotion detection, acoustic analysis, and word-use analysis can be effectively harnessed to assess palliative care quality in aspects of understanding, empathy, presence, emotion, and clarity. The ability of these technologies to automatically identify and analyze essential communication metrics underlines their potential in creating a foundation for feedback to clinicians in multiple types.

Our findings also emphasize the likelihood of varied performance across different communication features. Specifically, metrics such as understanding and emotion could demonstrate improved performance due to their intrinsic properties, straightforward implementation rules, or the capacity and adeptness of state-of-the-art NLP methods and tools in effectively capturing and analyzing these elements. This could be attributed to the algorithm's capacity to capture distinct markers associated with certain features or the relative difficulty in mimicking certain behaviors in the scripts. For instance, pauses and active listening, markers for `good' Presence, might be easier than to detect insufficient pauses and improper interruption, which indicate `bad' Presence. A meticulous analysis of these performance differentials from the perspective of NLP can guide further refinement of our algorithm.

Our study also shed light on the importance of extending the analytic focus beyond the linguistic and prosodic patterns of the provider. In addition to audio signals, wearable devices can measure physiological signals such as electrodermal activity and photoplethysmography thanks to embedded sensors. Non-verbal signals (body movement, facial expressions, etc.) can also be leveraged using ambient technology (e.g., cameras with motion capture software) installed in the patient's room. Moreover, if we incorporate the patient's behavioral signals as well (for applications where the patient's behavior is also in question and of importance), a number of analytical approaches will be applicable such as dynamical systems modeling and cross-correlation analysis to ascertain the interpersonal coordination processes manifested in different types of signals. We see these further developments as future steps of our work.

When comparing the CommSense to other existing technologies, it is crucial to highlight that the majority of these tools focus primarily on transcribing the content of the conversation or training \cite{ross2020story,zhang2022conversational}, extracting information from electronic health record \cite{ali2023using}, or understanding outcomes in a rule-based manner (e.g., developing a keyword library to detecting symptom talk \cite{durieux2023development}) \cite{bhatt2023use}. 
CommSense leverages a comprehensive computational framework to deeply analyze palliative care clinical conversations. Utilizing wearable sensing technology and sophisticated computational language models, it provides a nuanced assessment of conversational dynamics, which might be missed by human raters due to variability and biases. The system's metrics are supported by specialized language models or linguistic features for a comprehensive evaluation. Unlike traditional human rater-based methods, which are costly and not suited for real-time or longitudinal feedback, CommSense offers a scalable and efficient alternative, crucial for ongoing improvement in clinical communication.

Furthermore, the utility of wearable sensing technology in clinics, particularly smartwatches, is another important technical implication highlighted by our work. As a tool for data collection in this study, the smartwatch, coupled with a mobile sensing platform pilot-tested, proved instrumental. It reaffirmed the convenience and accessibility of wearable devices to integrate into healthcare environments without disrupting the clinical workflow and requiring additional human coders and burdens. This paves the way for future health communication technologies that transform captured data into actionable feedback for healthcare professions. 

These implications align with the broader trends in digital transformation in healthcare, including the adoption of artificial intelligence in clinical support \cite{dias2019artificial}, the utilization of sensing devices and IoT for clinical assistance \cite{haque2020illuminating}, and the increasing focus on addressing health disparities \cite{lyles2021focusing}. By showcasing how advanced technology can be utilized to facilitate enhanced patient-provider communication, CommSense positions itself at the frontier of this digital evolution, serving as a stepping stone for future technical developments in this crucial domain.

\subsection{Clinical Implications}

Patient-provider communication is crucial to both patient satisfaction and health outcomes. While we focused on palliative care and specifically evaluated outcomes related to the providers' communication skills in establishing rapport with the patient (e.g., exhibiting empathy) in this pilot study, our CommSense framework and metrics could be adapted to any number of healthcare interactions. 

The integration of digital tools like CommSense into healthcare settings holds promise for reshaping how we collect data on healthcare quality and transforming the landscape of opportunities to receive feedback and improve with relatively less burden compared to traditional clinical communication measurement. CommSense has three obvious clinical implications, including:

\begin{itemize}
    \item[1.] Education and training of clinicians: CommSense can facilitate objective evaluation of communication performance in  both simulated and real-world encounters to provide individualized and practical feedback, helping clinicians refine their communication skills.
    \item[2.] Quality improvement and continuing education for practicing clinicians: CommSense allows clinicians to assess and enhance their communication skills and organizational quality improvements after regular clinical training. Previously, this evaluation was only achievable via a clinic's patient panel or patient satisfaction surveys (e.g., the AIDET survey \cite{panchuay2023exploring}), which involve substantial labor and documentation burden.
    \item[3.] Clinical research: CommSense can help further delineate the most appropriate metrics to evaluate in patient-clinician communication and their impact on clinical outcomes.
\end{itemize}

Patient satisfaction is critical to assessing healthcare quality and equity, often gauged through traditional digital and paper surveys distributed post-interaction. CommSense can enhance existing measures by offering nuanced insights into clinician-patient communication dynamics and furnishing actionable feedback for improvement. While hurdles such as institutional resistance and technical complexities may surface, the potential advantages of adopting such tools significantly outweigh these challenges. As healthcare systems continue to evolve, the integration of digital tools like CommSense will play a crucial role in improving the quality of care and patient outcomes.

Our pilot study demonstrated that clinicians are highly interested in using (92.5\%) and receiving feedback (97.5\%) from CommSense, with a significant majority agreeing that such insights could enhance their awareness of the importance of their communication with patients (97.5\%), compassionate care skills (97.5\%), overall healthcare quality at organization level (97.5\%). This indicates that CommSense serves not only as a vital tool for continuing personal and professional development but also as a significant incentive for providers. Given the well-documented research linking suboptimal communication in healthcare to negative consequences for all involved \cite{thorne2008cancer,marshall2009teaching,vermeir2015communication}, CommSense, through providing feedback and insights in context, aims to mitigate these negative outcomes, yet the extent and specifics of its recommendations require further investigation in real-world clinics to understand its impacts. Additionally, these enhancements can help bridge the communication gap with underrepresented patients, who are at higher risk for suboptimal communication \cite{perry2018improving}, thereby potentially advancing healthcare equity.

This underscores CommSense's role as a crucial tool for professional development and a motivator for providers. Acknowledging existing literature that associates poor communication in healthcare with adverse outcomes, CommSense, through context-specific feedback, aims to address these issues, enhancing communication quality and exploring its recommendations' impacts and trade-offs in real-world settings. Furthermore, the system's potential to bridge communication gaps with underrepresented patients, who often experience suboptimal communication, could contribute to advancing healthcare equity, enabling clinicians to identify and correct biases in their interactions with diverse patient groups.

The negative impacts of poor communication are particularly felt by patients of minoritized groups \cite{anderson2009racial}. By providing objective communication performance to clinicians for all patient encounters, CommSense offers a path towards more equitable delivery of patient care. For example, with CommSense clinicians could see that they ask patients of color less frequently about the severity of their pain or are more likely to interrupt older patients - and then equipped with this new awareness, change their communication behavior accordingly.  Our `proof of concept' study revealed the majority of participants reported CommSense could help deliver more compassionate (i.e., equitable) healthcare.  In our future work in real patient care settings, we aim to more robustly evaluate this aspect of CommSense and its ability to reduce health disparities and enhance health equity.

In the wake of the COVID-19 pandemic, we have seen a marked increase in telehealth appointments via digital platforms such as Zoom. While telehealth has shown promise to help overcome barriers to patient care, we do not know the full impact on the patient–provider interaction.  A platform like CommSense could be adapted to better assess clinicians interactions in online platforms and provide actionable feedback to improve these virtual interactions. Another important area that CommSense could be applied is interprofessional education (IPE) between healthcare teams. Communication failure within healthcare teams is a leading cause of medical errors \cite{liang2000risks}, and IPE improves healthcare team communication and prepares students and clinicians to deliver safe high-quality team-based collaborative patient care \cite{cox2016measuring}. Our system and metrics could be adapted to help train medical teams communicate more efficiently and offer more automated assessment of providers interactions (e.g.; an attending and a resident). An additional example is leveraging a platform like CommSense to measure the therapeutic alliance or working alliance in mental healthcare (sometimes in social work as well) that quantifies a level of two-way engagement between the patient and the therapist. Current methods of evaluation use surveys and may benefit from the objective evaluation the CommSense framework offers when both parties' utterance is utilized.

Furthermore, CommSense, based on commercial ubiquitous smartwatches, offers a unique advantage for under-resourced hospitals, which may find traditional, communication training and evaluation methods cost-prohibitive. By providing a scalable and less resource-dependent alternative, CommSense has the potential to improve access to communication evaluation for clinicians who may practice in resource-limited settings, where access to quality communication training is hindered by geographical and/or financial barriers \cite{swartz2002integrating}.

\subsection{Ethical Concerns and Proposed Guidelines} \label{sec:ethics}

We identified three main categories of ethical concerns that may arise during the adoption of CommSense in the actual clinical arena (e.g., non-simulated space) and detailed them below. These concerns are in line with the generic ethical concerns for health monitoring or medical wearables but have distinctive considerations given the unique characteristics of CommSense and sensitive nature of recording conversations in the context of serious illness.

\subsubsection{\textbf{Informed Consent}} 

Given that CommSense records audio within the clinical environment, the approach to informed consent requires particularly careful attention. Healthcare providers and patients will sign written informed consent for the entire data collection period. Prior to each clinic visit, study personnel will confirm that consented patients, and any accompanying family members, agree to participate. We expect that this process will minimize burden, streamlining the consent process, for both clinicians and patients.

On the patient's side, special attention will be needed for marginalized groups to ensure that they are not pressured into consent and understand the technology and its implications. For instance, older adults or some patients suffering from chronic disorders may find it challenging to comprehend the technology in question, the types of data it captures, and/or the consent text itself \cite{segura2018ethical} to make an informed decision. Strategies such as simplified consent forms, multilingual support, and visual aids, in compliance with IRB regulations, may be introduced to accommodate the varied needs of these groups. Moreover, patients may perceive pressure from healthcare providers and simply agree to the monitoring; thus, it is critical that the informed consent process includes clear language that participation is completely voluntary and can be stopped at any time, without penalty. On the provider's side, there is a risk of perceived peer pressure or pressure from the management that they have to adopt this technology. The bottom line is that both patients and practitioners should have full agency over opting in or out of using CommSense and neither party should feel pressured. In our study, data collection is initiated only after obtaining informed consent from all involved parties, ensuring that patients and providers are fully aware of the nature of the data collection and its intended use, with their consent documented in a manner that is compliant and verifiable.

The other aspect to informed consent for both patients and providers is consenting to re-use or secondary use of the recorded audio and generated transcripts.  This is directly linked to the privacy issue shared by those participants surveyed, as these `raw' data contains sensitive patient information.  This aspect of informed consent is new for the healthcare domain with the advent of big data and will certainly become increasingly prevalent and necessary.  Healthcare institutions will need to adopt best practices from other industries and incorporate them into organizational policies and procedures.

\subsubsection{\textbf{Data Security}}
Data collected by CommSense is sensitive as it contains real-time spoken content potentially revealing personal information of the patient and the provider alike. As such, a potential leak and/or misuse of the collected data would have negative consequences. Measures must be taken to safeguard the security of data storage and processing. We securely transmit audio and sensor data captured during patient interactions to a processing server through an encrypted smartwatch application, protected by robust end-to-end encryption protocols. Upon reaching the secure server, data will undergo storage and processing. This step involves sophisticated algorithms in the proposed computational framework that featurize the data and transparent operationalization rules that identify and evaluate communication metrics automatically. Note, during this process, the data would be fully anonymized for both the patient and the provider, with particular care taken to ensure that sensitive information, such as test results, or protected health information (PHI), is not discernible.

Integrating CommSense's data security mechanisms into existing clinical information systems is feasible. By leveraging the existing security frameworks and collaborating with healthcare IT administrators, incorporating CommSense clinical conversation data and assessment into electronic health record (EHR) systems can be a practical long-term goal, ensuring its data is subject to the same security standards and ongoing review as other clinical data, as systems and compliance regulations change. In our pilot study using clinical simulations, data management was facilitated through Amazon AWS S3 servers with processing conducted offline. Future iterations will evolve towards a robust, real-time processing system that adheres to healthcare regulations such as HIPAA\footnote{Health Insurance Portability and Accountability Act, \url{https://www.hhs.gov/hipaa/index.html}} and GDPR\footnote{General Data Protection Regulation, \url{https://gdpr-info.eu}}, thus assuring the ethical management and confidentiality of sensitive clinical data.

\subsubsection{\textbf{Analytical Accuracy}} Tu et al. \cite{tu2021ethical} argues that inaccurate data collected by health monitoring wearable devices may result in incorrect health conclusions and unwanted or untimely interventions. We can only assume that with predictive models built with wearable-collected data as input, the output of CommSense will not always be true positive. When a prediction is not correct, it may trigger false feedback that may interfere with a practitioner's normal healthcare activity. To remedy this risk, we propose to be conservative in the generation of feedback and only do so when the confidence level of the predictive models is high. We also are committed to sharing feedback about communication performance that is contextualized to the clinical situation (for example, a high number of interruptions during an urgent interaction in the emergency department may be highly appropriate and should not be considered `negative' or `bad'). We will dedicate future work to experimenting with different strategies of issuing warnings to determine an optimal confidence level.

\subsubsection{\textbf{Healthcare Equity}}

As we prepare to implement CommSense in actual clinical settings on a wider scale, we must be mindful of assuring healthcare equity.  Marginalized and vulnerable patients face barriers in accessing healthcare that meets their needs, including palliative care.  They are at risk for gaps in care, such as healthcare providers missing opportunities to fully address their symptoms or take their wishes and goals of care into account.  A priority for CommSense will be capturing clinical conversations with marginalized patients in order to elucidate these gaps and inform strategies to mitigate them.  When seeking to enroll patients in future studies of CommSense, attention will be devoted to developing consent processes that are sensitive to potential barriers to study participation, such as health literacy or institutional distrust, so that marginalized groups can be well represented.  When transitioning from study contexts to real-world usual care contexts, consent processes will remain crucial for lowering barriers to patient participation.  Consent, data collection, and documentation will also need to be streamlined in ways that reduce pressure on overstretched staff and providers, particularly those who work with marginalized patients in often under-resourced clinical settings. Of note, we are actively engaging diverse stakeholder groups (administrators; clinicians; patients) to gain insights into potential barriers to future widescale CommSense implementation and strategies to ensure equitable, secure and meaningful integration in the clinical workflow.

\subsection{Limitations}

Despite progress in enhancing patient-provider communication, this work has several limitations which point to opportunities for future work:

Foremost among these limitations is the experimental setting for our study. Our research was carried out in simulated clinical scenarios with no real patient. As a result of this setting, we collected data by having forty participants enact eight scripts in a simulated clinic. In this context, the study also did not incorporate real patient feedback, which is a significant source of insight into the effectiveness and perception of empathy within the communication. The absence of this perspective leaves a gap in the comprehensive analysis of the interactions and the subsequent refinement of the CommSense system. Additionally, our preliminary feasibility test employed two distinct sets of scripts labeled as ``Good scripts'' and ``Bad scripts.'' However, real-world conversations typically blend positive and negative elements, adding complexity to the classification and measurement of communication metrics. Although our preliminary evaluation separately addressed these binary scenarios, CommSense is technically capable to manage such complexity, and future testing will prioritize mixed scenarios within naturalistic settings.

Furthermore, the source of the conversation scripts to train language models was limited, impacting our ability to fine-tune deep language models effectively. The general-purpose nature of these tools and datasets could potentially overlook the subtleties of empathetic exchanges which are central to patient care. Looking ahead, a vital step in this project will be to collect comprehensive data from actual clinical interactions and further test and refine the CommSense framework within real clinical contexts.

Additionally, the present approach predominantly relies on audio-based communication analytics which, though insightful, do not fully capture the interactions within the clinical context. Indeed, there are additional dimensions of communication which can be captured by smartwatch sensors and also are significantly relevant to quality patient-provider interactions, such as body language and conducive environment \cite{mondada2016challenges}.

\newcolumntype{s}{>{\hsize=.30\hsize}X}
\newcolumntype{t}{>{\hsize=.70\hsize}X}

{\renewcommand{\arraystretch}{1.5} 

\begin{table}[t!]
\caption{Roadmap for the Integration of CommSense in Clinical Practice}\label{tab:future}
\centering
\scalebox{0.95}{
{\small
\begin{tabularx}{\textwidth}{|s|t|}
\hline
\rowcolor{gray!30} 
\textbf{Milestone} & \textbf{Description} \\
\hline
\rowcolor{gray!10} 
Real-World Clinical Data Collection & We are collecting and analyzing clinical communication data in real-world settings. This will involve engaging with various healthcare settings to gather a diverse range of patient-provider interactions, which are crucial for testing the robustness and applicability of CommSense in various clinical contexts and patient demographics, enhancing the equitable delivery of patient care. \\
\hline
\rowcolor{gray!5}
Feedback from Relevant Stakeholders & We expand our feedback mechanism to include a more comprehensive set of stakeholders, such as patients, providers, administrators, and policy makers. This will ensure that CommSense is refined based on a holistic understanding of the needs and perspectives of all involved parties from both well-equipped and under-resourced organizations. \\
\hline
\rowcolor{gray!10}
Computational Test and Refinements & CommSense framework will undergo continuous testing and refinement to improve accuracy and reliability. This includes refining NLP algorithms and sensor data analysis, based on the feedback and data obtained from real-world clinical settings. Emerging technologies, including Generative Pre-trained Transformer (GPT) \cite{brown2020language}, will also be evaluated. \\
\hline
\rowcolor{gray!5}
Real-Time System Deployment & Designing and implementing a (nearly) real-time CommSense system, which involves NLP, secure edge and cloud computing, and wearable techniques is another key milestone. This would allow timely feedback to healthcare providers during and immediately after clinical interactions. \\
\hline
\rowcolor{gray!10}
Ethical and Regulatory Compliance & As we move towards wider deployment in actual clinical settings, adherence to ethical guidelines and regulatory compliance discussed in Section \ref{sec:ethics}, especially in collecting, transmitting and storing CommSense data and ensuring clear informed consent processes, is paramount. \\
\hline
\rowcolor{gray!5}
Engagement, Scalability, and Customization & A future step will engage healthcare professionals for training and integration of CommSense into clinical and educational practices. Particular focus will be on adapting the system to suit diverse healthcare environments and patient groups, promoting equitable and adaptable access and use. \\
\hline
\end{tabularx}}}
\end{table}
\renewcommand{\arraystretch}{1}} 

Other equity considerations include analytic challenges that could have more impact on vulnerable patients. For example, patients' cultural and language preferences may require use of an interpreter for proper communication with providers. This may challenge the capabilities of CommSense with multiple speakers and more than one language being used. As a result, the utility of CommSense may be diminished for this subgroup of patients. Patients who, for various reasons, are unable to speak for themselves are also vulnerable. Their chosen surrogate decision-makers may speak on their behalf, but this can significantly change the dynamics of communications with the providers. Emotional tone in these conversations may be different from more straightforward two-party interactions, and thus interpretation of CommSense output requires extra care to ensure analytic accuracy for all groups.

Lastly, the test in this feasibility study was preliminary, and not all aspects of the CommSense framework were fully tested or integrated into a real-time system. This limitation restricted the demonstration of the system's performance, such as real-time featurization and feedback, which were not available to participants during the study.

\section{Future Roadmap}

The next phase for CommSense involves working towards its implementation into actual clinical practice, evaluating its performance in real-world clinical contexts, and gathering acceptability data from a more diverse set of end-users. The key milestones and ongoing/proposed practices are outlined in Table \ref{tab:future}.

In summary, CommSense's roadmap plans to broaden its functionality and reach. Future efforts will concentrate on incorporating sophisticated computational technologies and expanding its use across diverse medical settings, tailoring it to real-world needs. We will refine its algorithms for deeper clinical communication analysis and integrate it into healthcare systems for comprehensive patient care. Additionally, CommSense is poised to improve healthcare policies and training programs, enhancing communication skills among professionals and trainees. The goal is to establish CommSense as a key healthcare communication tool, ultimately improving patient care quality and outcomes.
\section{Conclusion}

In this paper, we introduce CommSense, a wearable-enabled computational sensing framework designed to assess patient-provider communications by harnessing ubiquitous sensing and computing technology. Beginning with the identification of core communication metrics pertinent to quality clinical communications, we developed a suite of methods for audio and transcript data analysis and engineering. This allows for a more actionable measurement of patient-provider communication, touching upon key facets such as understanding, empathy, presence, emotion, and clarity. A pilot study involving N=40 participants, inclusive of licensed clinicians and nursing and medical students, evaluated the proposed computational framework in a controlled lab environment. The following takeaways emerge:

\begin{itemize}
    \item Feasibility: CommSense showed strong feasibility with commendable results in identifying communication metrics, achieving an average of 77.9\% in balanced accuracy, 71.2\% in precision, and 70.2\% in recall.
    \item Effectiveness: CommSense particularly excelled in the areas of understanding and emotion identification.
    \item Future Areas: There is room for improving the evaluation of presence and empathy metrics. Additional computational refinements and qualitative feedback will enhance its efficacy.
\end{itemize}

The present work offers valuable insights and lays the groundwork for the next steps in developing technologies and approaches for patient-provider communication in real-world clinical settings. We envision that CommSense will play a critical role in fostering better awareness and compassionate communication in healthcare, leading to improved quality and patient outcomes of palliative care.

\bibliographystyle{ACM-Reference-Format}
\bibliography{reference}

\end{document}